\documentclass[twocolumn,epjc3]{svjour3}          

\usepackage{amsmath}

\smartqed  

\RequirePackage{graphicx}
\RequirePackage{mathptmx}      
\RequirePackage{flushend}
\RequirePackage[numbers,sort&compress]{natbib}
\RequirePackage[colorlinks,citecolor=blue,urlcolor=blue,linkcolor=blue]{hyperref}

\journalname{Eur. Phys. J. C}

\begin{document}

\title{Distinguish the  $f(T)$ model from $\Lambda$CDM model with Gravitational Wave observations }


\author{Yi Zhang\thanksref{e1,addr1}
        \and
        Hongsheng Zhang\thanksref{addr2}
}

\thankstext{e1}{e-mail: zhangyia@cqupt.edu.cn}

\institute{College of  Science, Chongqing University of Posts and Telecommunications, Chongqing 400065, China \label{addr1}
          \and
          School of Physics and Technology, University of Jinan, West Road of Nan Xinzhuang 336, Jinan, Shandong 250022, China \label{addr2}
}


\maketitle

\begin{abstract}
 Separately, neither electromagnetic (EM) observations nor gravitational wave (GW) observations can distinguish between the $f(T)$ model and  the $\Lambda$CDM model effectively.  To break  this degeneration, we simulate the  GW measurement based on the coming observation facilities, explicitly the Einstein Telescope. We make cross-validations between the simulated GW data and factual EM data, including the Pantheon,  H(z), BAO and CMBR data, and the results show that they are consistent with each other. 
  Anyway, the EM data itself  have the $H_0$ tension problem which  plays critical role in the distinguishable problem as we will see.
 Our results show that the GW$+$BAO$+$CMBR data could  distinguish the $f(T)$ theory from the $\Lambda$CDM model in  $2\sigma$ regime.

\end{abstract}

\section{Introduction}\label{sec1}
The direct detection of gravitational wave (GW) confirms a major prediction of Einstein's General Relativity (GR)  and   initiates the era of gravitational wave physics\cite{Abbott:2016blz,Abbott:2016nmj,TheLIGOScientific:2016pea,Abbott:2017xzu,Abbott:2017vtc,Abbott:2017oio,Abbott:2017gyy,Abbott:2020niy,LIGOScientific:2021qlt}.  Anyway, not only General Relativity (GR)  which based on  the symmetric metric   with Levi-Civita connection could produce GW events, but also the  Teleparallel Gravity with Weitaenb$\ddot{o}$ck connection  \cite{Bengochea:2008gz,Linder:2010py,Wu:2010av,Cai:2015emx} could produce observable GW events.
 Until now,  56 GW events have been discovered.   In the coming decade, ground-based (e.g.Einstein Telescope (ET)  \cite{Punturo:2010zz,Sathyaprakash:2009xt} and space-based GW (e.g.Taiji \cite{Wu}, Tianqin \cite{Luo:2015ght}, and LISA \cite{Lisa}) experiments are predicted to discover more GW sources  and  provide a new and powerful tool to probe the fundamental properties of gravities.

 The most accurate observation on late acceleration, Planck data, favors $\Lambda$CDM model in GR \cite{Ade:2015xua,Ade}. As an extension of  Teleparallel Gravity, the $f(T)$ theory could provide the late acceleration for our universe as well.  The EM data can not distinguish  the $f(T)$ model   from the $\Lambda$CDM model  effectively \cite{Ferraro:2012wp,Clifton:2011jh,Zhang:2011qp,Awad:2017yod,Bamba:2010iw,Paliathanasis:2016vsw,Bamba:2010wb,Nassur:2016gzr,Salako:2013gka,Nesseris:2013jea,Nunes:2016qyp,Basilakos:2018arq,Nunes:2018xbm,Zhang:2012jsa,Capozziello:2018hly,Nunes:2018evm,Nunes:2019bjq,Qi:2017xzl,Yan:2019gbw} \footnote{Ref.\cite{Yan:2019gbw}   solves the $H_0$ tension problem in the frame of effective field theory including torsion which is different from our $f(T)$ models.  } which denotes the degeneration between  Teleparallel Gravity and General Relativity.
Furthermore,  the  Hubble constant  $H_0$ tension between the early  EM measurements (e.g. BAO and CMBR) and late EM measurement (e.g, Pantheon and H(z)),  which is the  most significant, long-lasting  and widely persisting tension,    is inherited in modified gravity.    Specifically, the typical cited prediction from Planck in a  flat $ \Lambda$CDM model for the Hubble constant is $H_0=67.27\pm0.60 km/s/Mpc$ at $68\%$ confidence level (CL)  for Planck 2018  \cite{Aghanim:2018eyx}, while that from the SH0ES Team is $H_0=73.2\pm 1.3 km/s/Mpc$ \cite{Riess:2020fzl}  which yields a  $4.14\sigma$ tension (see e.g.\cite{DiValentino:2021izs} for a review and references therein for various models to solve the
$H_0$ tension proposed so far).

For GW signals, besides the $``+"$ and $``\times"$ polarization patterns,  the simplest $f(T)$ gravity in four dimensional space-time provides one extra degrees of freedom, namely a massive vector field.
    Such an extra degree   of tensor perturbation in $f(T)$ Teleparallel Gravity  does not propagate because of its Yukawa-like potential. Then, the effective degree of tensor perturbation in $f(T)$ model could be regarded as  the same as GR in the Post-Minkowskian limit \cite{Bamba:2013ooa,Cai:2018rzd,Abedi:2017jqx}.  The literature \cite{Basilakos:2018arq}  constrains the $f(T)$ model by using  the GW phase effect based on the TaylorF2 GW waveform.      It shows that detection sensitivity  within  ET can improve  up  two orders of magnitude of  the current bound on the $f(T)$ gravity. The TylorF2  form is based on post-Newtonian (PN) method  which is related to metric gravity, and uses the stationary phase approximation which assumes slowly varying amplitude and phase. Then,  it is  necessary to consider the effect of  the GW amplitude  which is proportional to  the luminosity distance \cite{Baker:2017hug,Ezquiaga:2017ekz}.  The GW sources are  distance indicators which is called dark siren as shown in Refs. \cite{Holz:2005df,Schutz}.
 Furthermore,  the propagation of  gravitational wave is different in modified gravity and  General Relativity \cite{Belgacem:2019pkk,Amendola:2017ovw,Belgacem:2017ihm,Lagos:2019kds,Calcagni:2019kzo}.

Numerical simulations present valuable approach to forecast results of surveys and targeted observations that will be performed with next generation instrument like Einstein Telescope \cite{gwbook,Zhao:2010sz,Cai:2016sby}.    The Einstein Telescope will detect thousands of NSB (Neutron Star Binary) and BHB (Black Hole Binary) mergers to probe the cosmic expansion at high redshifts.
 In view that only the EM data cannot distinguish the $f(T)$ model from the $\Lambda$CDM model effectively, we will simulate the GW data from  Einstein Telescope design and the $f(T)$  model parameters.
    And we try to  search the $H_0$ tension and distinguishable problem in $f(T)$ theory by using the GW data.

   The outline of the paper is as follows. In  Section \ref{ft}, we introduce  the  Teleparallel Gravity  and two explicit $f(T)$ models ($f_p$CDM and $f_e$CDM models).
   In Section \ref{lum}, we display the differences between EM and GW data.
In Section \ref{data}, we introduce our EM data firstly, and then our simulated GW data. After checking consistence, we  combine  the GW data and the EM data  to constrain the $f(T)$ models separately. In Section \ref{result}, we discuss the constraining results. At last, in Section \ref{sum},  we concisely summarize this paper.


    \section{A brief review of $f(T)$ theory}\label{ft}
In  $f(T)$ cosmology, the action reads,
  \begin{eqnarray}
  \nonumber
  S&=&\int d^{4}x \sqrt{-g}(f(T)+\Omega_{m0}(1+a)^3)\\
  &=& \int d^{4}x \sqrt{-g}(T+F(T)+\Omega_{m0}(1+a)^3),
  \end{eqnarray}
   where $ \sqrt{-g}$ is the determination of co-tetrad $e_{\mu}^{A}$  in $f(T)$ model,  $T$ is the torsion scalar playing the role of $R$ in GR, $\Omega_{m0}$ stand for the present matter  energy density, the index $``0"$ denotes the present value, the $F(T)$ term plays the role of acceleration. Here, we neglect the effects of radiation for the evolution of
  the universe.   And, we assume the background manifold to be a spatially flat Friedmann-Robert-Walker (FRW) universe, then
  the torsion scalar reads $T=-6H^{2}$ where $H$ is the Hubble parameter. Routinely, one obtains  effective energy density $\rho_{T}$ and  effective pressure $p_{T}$ for the FRW universe,
  \begin{eqnarray}
  && \rho_{T}=\frac{1}{16\pi G}(-12H^{2}f_T-f+6H^{2}),\\
  &&p_{T}=\frac{f-Tf_{T}+2T^{2}f_{TT}}{16\pi G(1+f_T+2Tf_{TT})},
  \end{eqnarray}
  where $f_T=df/dT$, $f_{TT}=d^{2}f/dT^{2}$.
        For convenience, we define  a dimensionless parameter related to cosmological model,
  \begin{eqnarray}
  E^{2}(z)=\frac{H^2}{H_0^2}=\Omega_{m0}(1+z)^{3}+(1-\Omega_{m0})\frac{f-2Tf_T}{T_0\Omega_{T0}},
    \end{eqnarray}
  where   $\Omega_{T0}=8\pi G \rho_{T0}/(3H_0^{2})$.
  And, the effective equation of state (EoS) parameter reads
  \begin{eqnarray}
  w_{eff}=\frac{p_{T}}{\rho_{T}}=-\frac{f/T-f_T+2Tf_{TT}}{[1+f_T+2Tf_{TT}][f/T-2f_T]}.
  \end{eqnarray}
 Once $F_T= constant$, then $w_{eff}=-1$, the model comes back to the $\Lambda$CDM model.
In the following text, we explore one of the most interesting and tractable $f(T)$ models with one extra parameter.

\subsection{The power-law form: $f_{p}$CDM model}
 In the literatures, the power-law $f(T)$ model \cite{Bengochea:2008gz} (hereafter $f_{p}$CDM model)   is an interesting and notable model,
\begin{eqnarray}
F(T)=\alpha(-T)^{b},
\end{eqnarray}
where $ \alpha=(6H_0^{2})^{1-b}(1-\Omega_{m0})/(2b-1)$.  Essentially the distortion parameter $b$ is the solo new freedom which quantifies deviation from the $\Lambda$CDM model.
 When $b=0$, this model degenerates to the $\Lambda$CDM model.

\subsection{The square-root exponential form: $f_{e}$CDM model}
Then, we introduce the square-root exponential model   \cite{Linder:2010py}  (henceafter $f_{e}$CDM model )
\begin{eqnarray}
F(T)=\alpha T_0(1-e^{-p\sqrt{T/T_0}}),
\end{eqnarray}
where $\alpha=(1-\Omega_{m0})(1-(1+p)e^{-p}) $ and   $p$  is a model parameter.   A similar model is proposed in literature, in which  $f(T)=\alpha T_0(1-e^{-pT/T_0})$ \cite{Nesseris:2013jea}.
 For convenience,  we set $b=1/p$.  Anyway, $b\to+0$  corresponds to $p\to +\infty$, while $b\to-0$  corresponds to $p\to-\infty$.
Then, for us, getting  across $b=0$ means crossing the singularity $p$. Therefore,  in the numerical processes, for convenience, we set  the prior $b>0$ which is favored by Ref \cite{Basilakos:2018arq}.    When $b<0$, $e^{-p\sqrt{T/T_0}}>1$  grows exponentially. In numerical calculation, since it is difficult to cross $b=0$ we just set the prior that $b>0$.

\subsection{A short Discussion}
    In both $f(T)$  models,  an additional parameter $b$ appears.  When $b=0$,   the $f(T)$ model comes back to $\Lambda$CDM model.  And  $b\neq 0$ in $f(T)$ model indicates an essential deviation from  $\Lambda$CDM model.

  \section{The luminosity distance in EM data and GW data}\label{lum}
 The    gravitational wave (GW) standard sirens  offer a new independent way to probe the cosmic expansion. From the GW signal, we can measure the luminosity distance $d_{L}^{GW}$ directly, without invoking the cosmic distance ladder, since the standard sirens are self-calibrating.  The gravitational waves from compact systems are viewed as standard sirens to probe the evolution of the universe \cite{Holz:2005df,Schutz}.   We can extract luminosity distance from the GW amplitude
  \begin{eqnarray}
  \label{dynamicalhA}
   h_A=\frac{4}{d_L^{GW}} (\frac{GM_c}{c^2})^{5/3}(\frac{\pi f_{GW}}{c})^{2/3},
\end{eqnarray}
 where $ h_A$ is the GW amplitude, ``$A$" could be ``$+$" or ``$\times$", $d_L^{GW}$ is the luminosity for  gravitational wave, $M_c$ is the chirp mass, and $f_{GW}$ is the GW frequency. Here we ignore the lower index of ``$A$" because the ``$+$"  or ``$\times$"  polarization pattern of GW shares the same luminosity form.
 From Eq.(\ref{dynamicalhA}), one sees the significant property of GWs, that is, the amplitude of GWs is inversely proportional to its luminosity distance. If the EM counterpart of the GW event is observed, a redshift measurement of the source enables us to constrain the cosmic expansion history.
     For the $f(T)$ theory, the evolution equation of GW in Fourier form is  \cite{Bamba:2013ooa,Cai:2018rzd,Abedi:2017jqx},
 \begin{eqnarray}
 \label{fTgw}
 \ddot{h}_{k}+3H(1-\beta_{T})\dot{h}_{k}+\frac{k^2}{a^2}h_{k}=0,
 \end{eqnarray}
 where  the extra  friction term reads,
 \begin{eqnarray}
 \label{betaT}
\beta_T=-\frac{\dot{f_T}}{3Hf_T}=-\frac{2\dot{H}}{3H^2}\frac{Tf_{TT}}{f_T}.
\end{eqnarray}
When $\beta_T=0$, it reduces to the  $\Lambda$CDM Model in GR gravity.  When $f_{TT}=0$, it means $F(T)=c_1T+c_2$ where $c_1$ and $c_2$ are constants, as the $c_1$ will be nomorlization by the T, so $F(T)=constant$, the model comes back to the $\Lambda$CDM model.
The extra friction term ($3H\beta_{T}\dot{h}_{ij}$) affects the amplitude of $h_{ij}$. If  $\beta_T>0$, the damping will be slower. If  $\beta_T<0$, the damping will be more remarkable.
 Then, if the cosmic evolution is deviated from $\Lambda$CDM model, the constraint result of  EM data should reveal a non-zero $\beta_T$ as well.

To simplify the propagation equation of  of $h_A$ in $f(T)$ theory, we define a new scale factor $\tilde{a}$ as
 $ \tilde{a}'/\tilde{a}=\mathcal{H}[1-\beta_T]$,
  where $\mathcal{H}=a'/a$ and the prime ($``'"$) is respect to the comoving time. And, after defining a new parameter $\chi_{k}=h_k/\tilde{a}$,
the propagation equation of  of $h_A$ becomes,
\begin{eqnarray}
\chi^{''}_k+(k^2-\frac{\tilde{a}''}{\tilde{a}})=0.
\end{eqnarray}
 When the gravitational  wave propagates across cosmological distance,  $\chi_k$ decreases as $1/\tilde{a}$ rather than $1/a$.
In small scale, the term $\tilde{a}''/\tilde{a}$ is  negelectable.  Then, this effect is best to test in cosmological scale.

 As shown in Ref.\cite{Belgacem:2017ihm,Lagos:2019kds}, the relation between the EM luminosity distance and GW luminosity distance is
\begin{eqnarray}
d_{L}^{GW}(z)=\frac{a(z)}{\tilde{a}(z)}d_{L}^{EM}=d_{L}^{EM}exp{(-\int_{0}^{z}\frac{dz'}{1+z'}\beta_T(z'))},
\label{dLgw}
\end{eqnarray}
where $d_{L}^{EM}=(1+z)/H_0\int_{0}^{z}d\tilde{z}/E(\tilde{z})$.
This equation shows that the extra friction  term ($3H\beta_{T}\dot{h}_{ij}$) makes  the EM luminosity data are different from that of GW. Meanwhile, the difficulty of test of luminosity of SN Ia roots in distance, the test of luminosity of GW is redshift. The two observational difficulties are complementary, which can be solved in the meantime by a combined
constraint using EM and GW data.
The joint EM and GW data are predicted to  yield a  non-zero $\beta_T$ and break the parameter degeneration \footnote{The two sets of data (EM and GW) both indicate $\beta_T \neq 0$ as shown in Table \ref{tab}, which denotes an acceleration expansion as well.}.


\section{ The Data}\label{data}

First, we constrain cosmological models by using realistic EM data and simulated GW data.   The EM data contain the  Pantheon, H(z), BAO and  CMBR data which are widely used in explorations of cosmology.
 We follow the simulation in Ref.\cite{Cai:2016sby} to explore the cosmological constraints on $f(T)$ or General Relativity by simulated data based on Einstein Telescope and the $f(T)$ model. If the simulated GW data are consistent with the EM data, it is reasonable to combine them to constrain cosmologies.

  \subsection{The EM data}
  Here, we briefly introduce the Pantheon, H(z), BAO and  CMBR data.

 The Pantheon  sample of  1048 supernovae Ia (SNe Ia), whose  redshift range is $0.01 < z <2.3$,  combines the subset of 276 new Pan-STARRS1 (PS1), SNe Ia with useful distance estimates of SNe Ia from SNLS, SDSS, low-z and Hubble space telescope (HST) samples \cite{Scolnic:2017caz} .

   In the  H(z) measurements, $41$ data  \cite{Zhang:2012mp,Jimenez:2003iv,Moresco:2012by,Moresco:2015cya,Gaztanaga:2008xz,Simon:2004tf,Chuang:2011fy, Xu:2012fw,Moresco:2016mzx,Blake:2012pj,Stern:2009ep,Samushia:2012iq,Busca:2012bu,Font-Ribera:2013wce,Delubac:2014aqe} are considered which could be obtained via two ways. One is to calculate the differential ages of passively evolving galaxies, usually called cosmic chronometer. The other is based on the detection of radiation BAO features.

The property of baryon acoustic oscillation (BAO) in the clustering of matter in the universe serves as a robust standard ruler and hence can be used to map the expansion history of the universe.
 For the BAO data, the ratios of distances and the so called dilation scale
$D_V(z)$ at different redshifts $z$ are taken after
\cite{Percival:2009xn,Blake:2011en,Beutler:2011hx}
\footnote{The BAO and $H(z)$  measurements used in this paper are summarized in Table 1 and 2 of Ref \cite{Qi:2017xzl}.}.

  The $R$ parameter, which extracts the information in cosmic microwave background (CMB),
 is  defined as
 \begin{eqnarray}
 R=\sqrt{\Omega_{m0}}\int_0^{z_{*}}dz'/E(z'),
 \end{eqnarray}
  where $z_{*}=1090.43$
 denotes the decoupling redshift.
  We use the first year data of Planck which show  $R=1.7499\pm 0.0088$ \cite{Ade}.

Roughly speaking, we regard the Pantheon, H(z) data as the late EM  measurement, and the BAO and  CMBR data as the early one.

\subsection{The GW data simulation}
 The detected GW events are  limited, however the upcoming apparatuses for GWs are expected to detect  unprecedented amount of GW events.  It is an urgent need to simulate the GW data for the upcoming apparatuses to
 display the expected improvements in science.    We  simulated GW data whose  $d_L-z$ relation depends on the ET design and  the cosmological model parameters $\Omega_{m0}$, $H_0$ and $b$.
  We divide the whole process into two steps.  First, we  constrain the $f(T)$ models by using the EM data. Then, by using the best fitted  value of $f(T)$ models as the fiducial value of the GW simulation, we calculate the GW
  luminosity distance based on Eq.(\ref{dLgw}) as the fiducial  value.
    The simulation is  parallel the well-known simulation for the case of $\Lambda$CDM model \cite{Cai:2016sby}.

The gravitational waves originate from binary neutron stars or black holes.
   In the transverse traceless (TT) gauge, the amplitude $h(t)$ can be obtained by detector
  \begin{eqnarray}
  h(t)=F_{+}(\theta,\phi,\psi) h_{+}(t)+F_{\times}(\theta,\phi,\psi) h_{\times}(t),
 \end{eqnarray}
where $F_{+}$, $F_{\times}$ are the antenna pattern functions sensed by the detector, $\psi$ is the polarization angle, and $(\theta,\phi)$ are angles describing the location of the source in the sky relative to the detector.
Base on the sensitivity  of the Einstein Telescope we study  the constraint precision on the cosmological parameters by simulating binary systems of NS-NS and NS-BH that have an accompanying EM signal by using Fisher matrix \cite{Cai:2016sby} which is expected 1000 events will be found in the Einstein Telescope in one year. It could be event of NS-BH, or NS-NS. Roughly we assume there are 500 NS-BH events vs 500 NS-NS events.   The redshift distribution of the events is followed by Ref. \cite{gwbook,Zhao:2010sz,Cai:2016sby}.    The  simulation redshift is chosen as  $z<5$.
The performance of a GW detector is characterized by its one-side noise power spectral density  (PSD). The PSD of   Einstein Telescope is taken to get our error of  simulating data as shown in Ref.\cite{Zhao:2010sz}.
Furthermore, the luminosity distance is also affected by an additional error lens due to the weak lensing effects. According to the studies
in \cite{gwbook,Zhao:2010sz}, we assume  the error is $ 0.05z$. Thus, the total
uncertainty on the measurement of $d_L$ is taken to be
\begin{equation}
\label{error}
\sigma_{d_L}=\sqrt{\sigma_{inst}^2+\sigma_{lens}^2}=\sqrt{(\frac{2d_{L}}{\rho})^2+(0.05z d_L)^2},
\end{equation}
where $\rho$ is the ratio of signal to noise which  is usually chosen as   $\rho>8$.      And based on the above discussions , we derive the GW simulated data.



\begin{figure*}  \centering
{ \includegraphics[width=8cm,height=8cm]{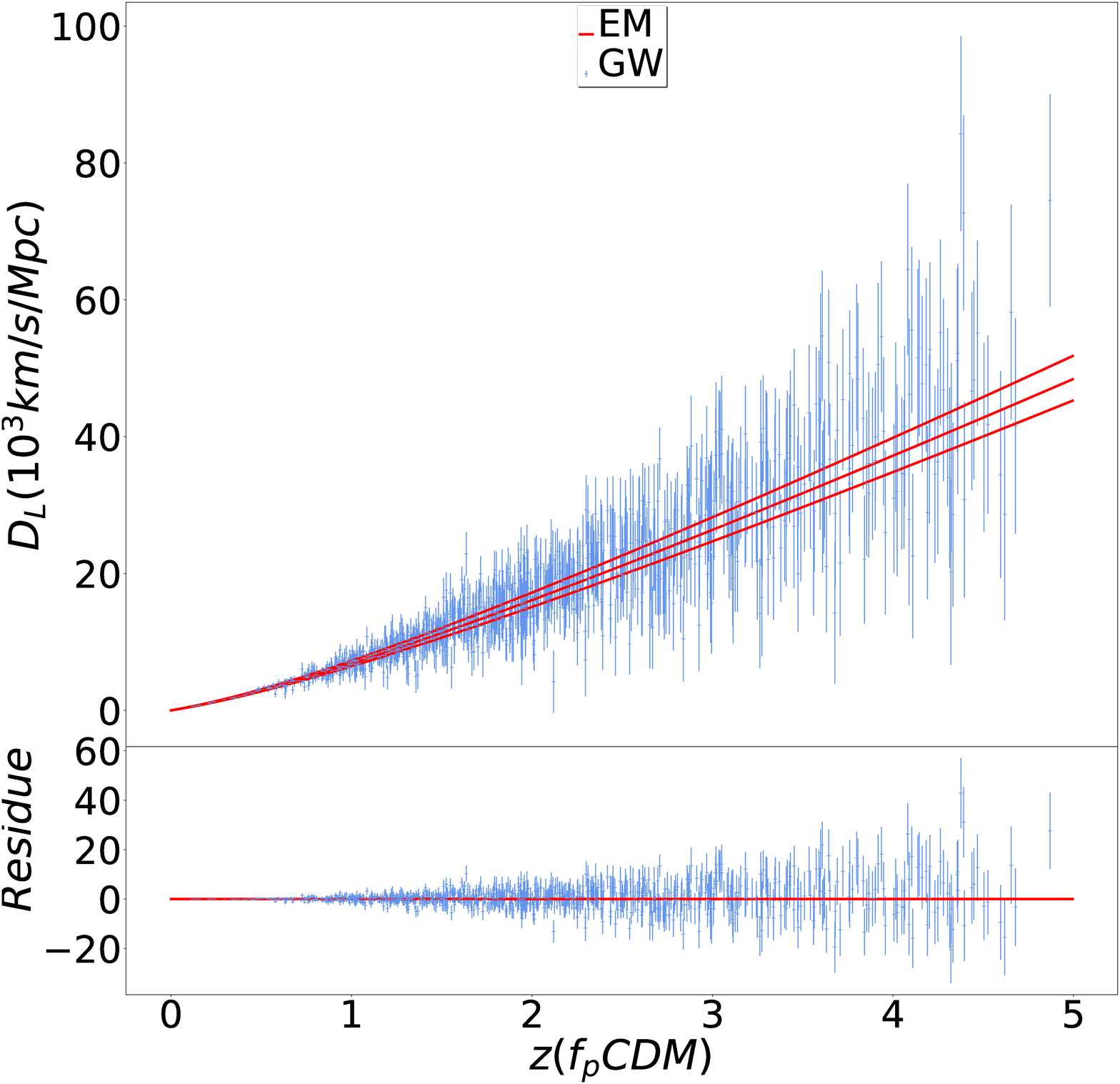}}
{ \includegraphics[width=8cm,height=8cm]{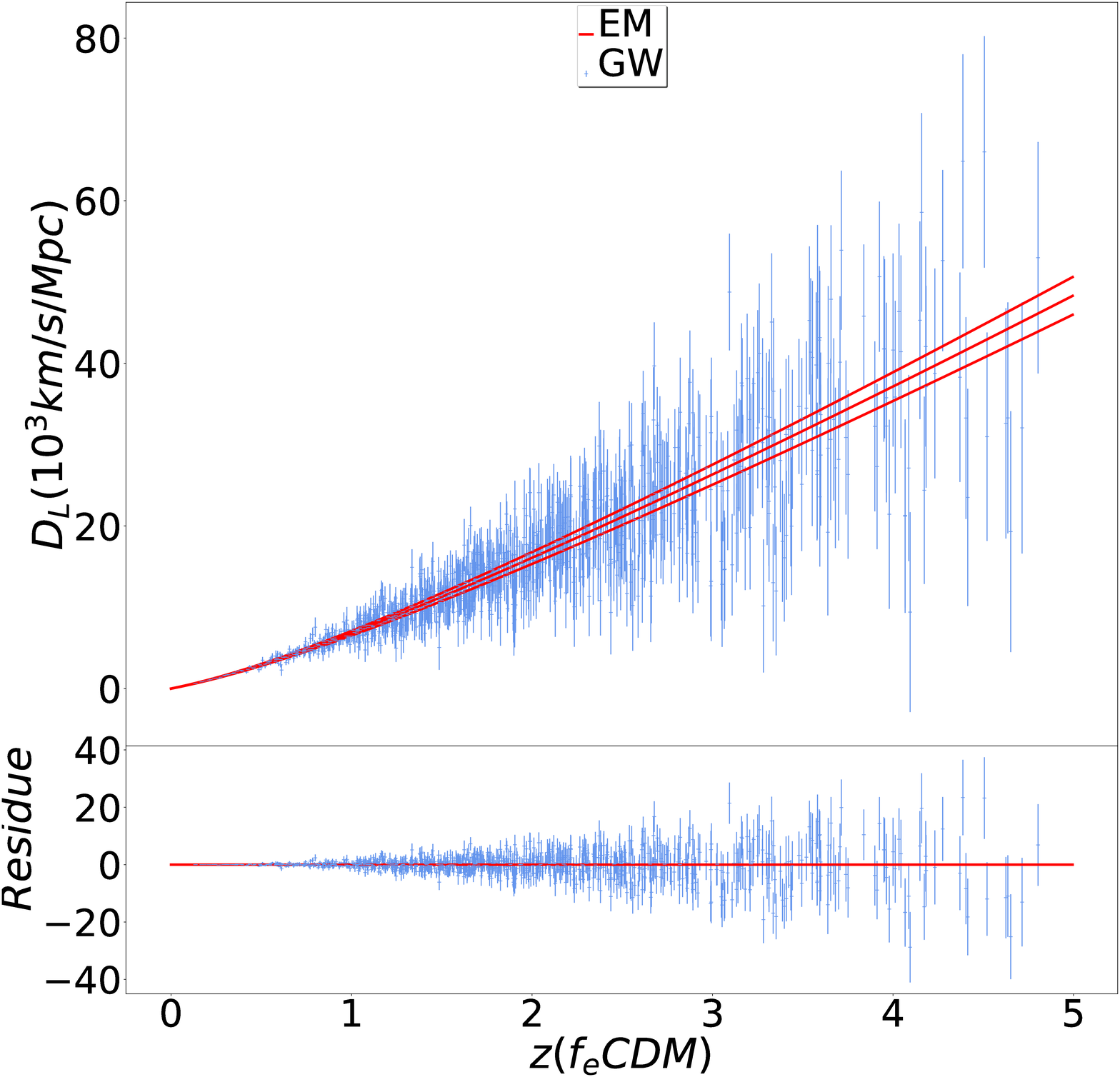}}

  \caption{   The data with error-bar are  the simulated GW data. The  red lines are the best fitted values and the $1\sigma$ errors  from the EM fitting results. Specifically, the left panel is for the $f_p$CDM model, while the right panel is for the $f_e$CDM model.}
  \label{dl}
\end{figure*}

\begin{table*}
\caption{Best fitted values with $1\sigma$ and $2\sigma$ standard errors for the   GW, EW and GW $+$ EM  (combined) data for the $f_e$CDM and  $f_p$CDM models.}
\label{tab}
\begin{tabular*}{\textwidth}{@{\extracolsep{\fill}}llllllll@{}}

	 \hline  $f_pCDM$  &  $\Omega_{m0}$ &  $H_0(km/s/Mpc)$  &  $ b$ &$\beta_T$& $w_{eff}$  \\	
	 \hline
	
	GW    & $ 0.263^{+0.048+0.068}_{-0.021-0.089}$
		     		     &$ 68.21^{+1.32+2.51}_{-1.32-2.61}$
		     &$ -0.122^{+0.624+0.868}_{-0.361-1.023}$
		
		     &$ -0.018^{+0.131+0.152}_{-0.032-0.241}$
		     &$ -1.003^{+0.095+0.239}_{-0.146-0.212}$
		        & $2.69\sigma$ \\

EM &  $0.293^{+0.010+0.020}_{-0.010-0.019}$
   &$ 68.41^{+1.23+2.44}_{-1.24-2.41}$
 &$ -0.071^{+0.095+0.171}_{-0.078-0.182}$
 &$ 0.012^{+0.017+0.031}_{-0.015-0.033}$
 &$-1.018^{+0.024+0.047}_{-0.024-0.046} $& $2.67\sigma$
 \\

		     	GW$+$Pantheon  & $0.197^{+0.047+0.075}_{-0.038-0.083}$
	& $ 67.62^{+0.50+0.99}_{-0.50-0.97}$
&$ 0.441^{+0.240+0.349}_{-0.151-0.382}$ &
 $-0.151^{+0.120+0.171}_{-0.061-0.196}$&
 $-0.861^{+0.070+0.131}_{-0.070-0.131}$ & $4.01\sigma$
\\

GW$+$H(z)&$ 0.262^{+0.020+0.034}_{-0.036-0.014}$&
$68.20^{+0.65+1.31}_{-0.65-1.31}$&
 $0.031^{+0.219+0.395}_{-0.178-0.410}$&
$-0.014^{+0.052+0.081}_{-0.031-0.092}$&
$-0.987^{+0.050+0.110}_{-0.061-0.110}$& $3.46\sigma$\\

GW$+$BAO& $0.274^{+0.016+0.031}_{-0.016-0.032}$&
 $68.40^{+0.63+1.21}_{-0.63-1.21}$&
$-0.161^{+0.261+0.412}_{-0.170-0.458}$&
 $0.022^{+0.038+0.066}_{-0.030-0.070}$&
 $-1.034^{+0.051+0.099}_{-0.051-0.101}$&$3.30\sigma$ \\

GW$+$CMBR&
$0.277^{+0.009+0.019}_{-0.009-0.018}$&
 $68.61^{+0.57+1.12}_{-0.57-1.12}$&
 $-0.226^{+0.119+0.203}_{-0.090-0.218}$&
$0.036^{+0.014+0.028}_{-0.014-0.029}$&
$-1.052^{+0.022+0.043}_{-0.022-0.044}$   &$3.21\sigma$\\

 GW$+$Pantheon$+$H(z) &$0.258^{+0.020+0.035}_{-0.016-0.037}$&
 $67.77^{+0.50+0.98}_{-0.50-0.98}$&
$0.141^{+0.159+0.292}_{-0.141-0.310}$&
$-0.037^{+0.047+0.074}_{-0.030-0.083}$&
 $-0.956^{+0.043+0.097}_{-0.088-0.050}$ &$3.90\sigma$ \\

GW$+$Pantheon$+$BAO &$0.269^{+0.013+0.027}_{-0.013-0.026}$&
$67.82^{+0.46+0.90}_{-0.46-0.90}$&
 $0.042^{+0.110+0.211}_{-0.110-0.279}$&
$-0.009^{+0.018+0.050}_{-0.014-0.053}$&
 $-0.989^{+0.020+0.065}_{-0.072-0.024}$ &$3.90\sigma$
\\

GW$+$BAO$+$CMBR&$0.280^{+0.009+0.017}_{-0.009-0.016}$&
 $68.51^{+0.57+1.12}_{-0.57-1.12}$&
 $-0.223^{+0.118+0.221}_{-0.099-0.221}$&
 $0.035^{+0.016+0.029}_{-0.014-0.032}$&
 $-1.052^{+0.023+0.048}_{-0.023-0.045}$& $3.28\sigma$ \\

		   GW $+$EM  &  $0.286^{+ 0.007+0.015}_{- 0.007-0.014}$
		   		   &$ 68.06^{+0.46+0.90}_{-0.46-0.89}$
		   &$ -0.117^{+0.086+0.152}_{-0.070-0.160}$
		
		   &$ 0.020^{+0.013+0.025}_{-0.013-0.026}$
		   &$ -1.030^{+0.019+0.038}_{-0.019-0.038}$
		     & $3.73\sigma$   \\

\hline  $f_eCDM$ &  $\Omega_{m0}$ &  $H_0(km/s/Mpc)$ &    $b$& $\beta_T$&$w_{eff}$ & Tension    \\
		\hline

	GW&
$0.261^{+0.028+0.040}_{-0.015-0.050}$&
 $67.46^{+0.78+1.33}_{-0.64-1.38}$&
$b < 0.404< 0.708$&
 $-0.249^{+0.235+0.251}_{+0.046-0.41}$&
 $-0.869^{+0.034+0.231}_{-0.125-0.154}$&  $3.88\sigma$ \\	
		
		EM  &  $0.296^{+0.010+0.020}_{-0.010-0.019}$
		  &$67.71^{+1.11+2.10}_{-1.11-2.09}$&$b < 0.162< 0.251$&$ -0.031^{+0.028+0.031}_{-0.013-0.042}$ &$-0.959^{+0.014+0.050}_{-0.041-0.042}$ &$3.21\sigma$  \\

GW$+$Pantheon &$0.269^{+0.023+0.035}_{-0.014-0.041}$&
 $67.58^{+0.50+0.97}_{-0.50-0.99}$&
 $0.261^{+0.154+0.261}_{-0.168-0.261}$&
 $-0.093^{+0.072+0.090}_{-0.041-0.169}$&
 $-0.906^{+0.032+0.11}_{-0.091-0.095}$& $4.03\sigma$\\

GW$+$H(z)&$0.276^{+0.011+0.021}_{-0.011-0.023}$&
 $67.68^{+0.71+1.11}_{-0.52-1.27}$&
$b < 0.282< 0.432$&
 $-0.061^{+0.412+0.750}_{-0.411-0.644}$&
$-0.923^{+0.022+0.092}_{-0.075-0.077}$& $3.84\sigma$\\

GW$+$BAO& $0.280^{+0.012+0.023}_{-0.012-0.023}$&
 $67.67^{+0.60+1.03}_{-0.47-1.08}$&
$b < 0.225< 0.372$&
 $-0.048^{+0.063+0.121}_{-0.148-0.077}$&
$-0.941^{+0.017+0.080}_{-0.059-0.060}$   & $3.93\sigma$\\

GW$+$CMBR&$0.292^{+0.010+0.019}_{-0.010-0.018}$&
 $67.55^{+0.43+0.82}_{-0.43-0.88}$&
$b < 0.145< 0.233$&$ -0.028^{+0.027+0.028}_{-0.009-0.036}$&
$ -0.963^{+0.012+0.046}_{-0.036-0.037}$  & $4.13\sigma$  \\

GW$+$Pantheon$+$H(z)&$0.278^{+0.011+0.020}_{-0.010-0.023}$&
$67.70^{+0.51+0.92}_{-0.46-0.99}$&
$0.211^{+0.128+0.168}_{-0.114-0.211}$&
$-0.058^{+0.056+0.058}_{-0.024-0.065}$&
$-0.929^{+0.041+0.069}_{-0.047-0.071}$&$3.96\sigma$\\

GW$+$Pantheon$+$BAO& $0.282^{+0.012+0.021}_{-0.010-0.023}$&
$67.64^{+0.45+0.86}_{-0.45-0.92}$&
$b < 0.220 < 0.338$&
$-0.045^{+0.044+0.045}_{-0.013-0.058}$&
$-0.944^{+0.020+0.065}_{-0.056-0.057}$& $4.04\sigma$ \\

GW$+$BAO$+$CMBR& $0.292^{+0.009+0.018}_{-0.009-0.017}$&
$67.54^{+0.42+0.78}_{-0.38-0.84}$&
$b < 0.142< 0.230$&
$-0.027^{+0.027+0.027}_{-0.008-0.036}$&
$-0.964^{+0.011+0.045}_{-0.035-0.036}$  & $4.16\sigma$ \\

		  GW $+$ EM  &  $ 0.294^{+0.007+0.015}_{-0.007-0.014} $
		   &$ 67.84^{+0.36+0.69}_{-0.36-0.74} $
		  &$b < 0.154< 0.236 $
		  & $-0.029^{+0.026+0.029}_{-0.012-0.038} $  &$ -0.961^{+0.012+0.045}_{-0.038-0.039} $
		     & $3.97\sigma$ \\

           \hline		
\end{tabular*}
\end{table*}


      \begin{figure*}  
      \centering

{ \includegraphics[width=17cm,height=14cm]{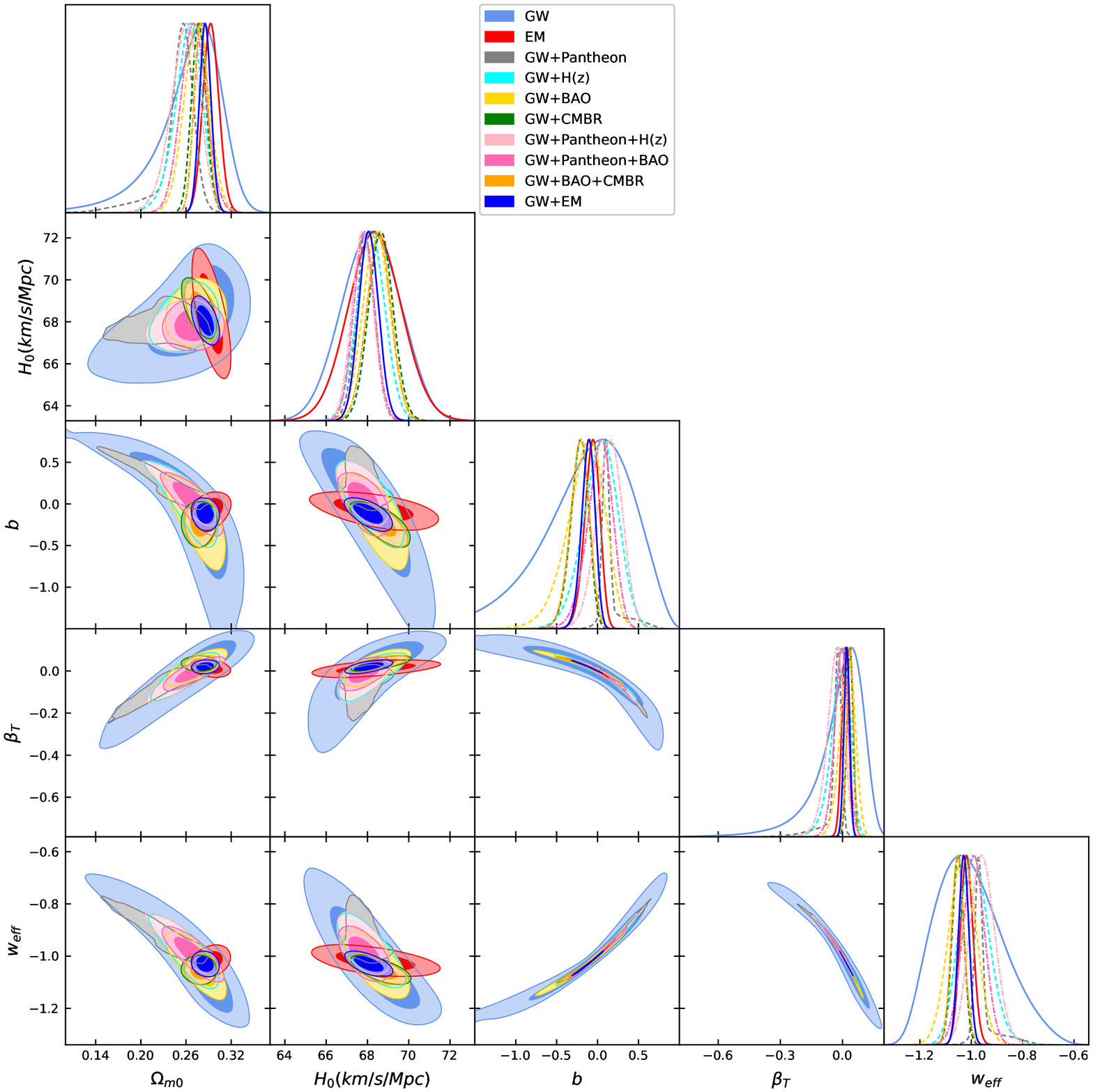}}

  \caption{ The probability density functions (pdfs) ,   the $68\%$  ($1\sigma$) and $95\%$ ($2\sigma$) confidence regions for the parameters  ($\Omega_{m0}$, $H_0$, $b$, $\beta_T$ and $w_eff$) in the $f_p$CDM model respectively.    }
  \label{ft1tri}
\end{figure*}

     \begin{figure*}  
      \centering

{ \includegraphics[width=17cm,height=14cm]{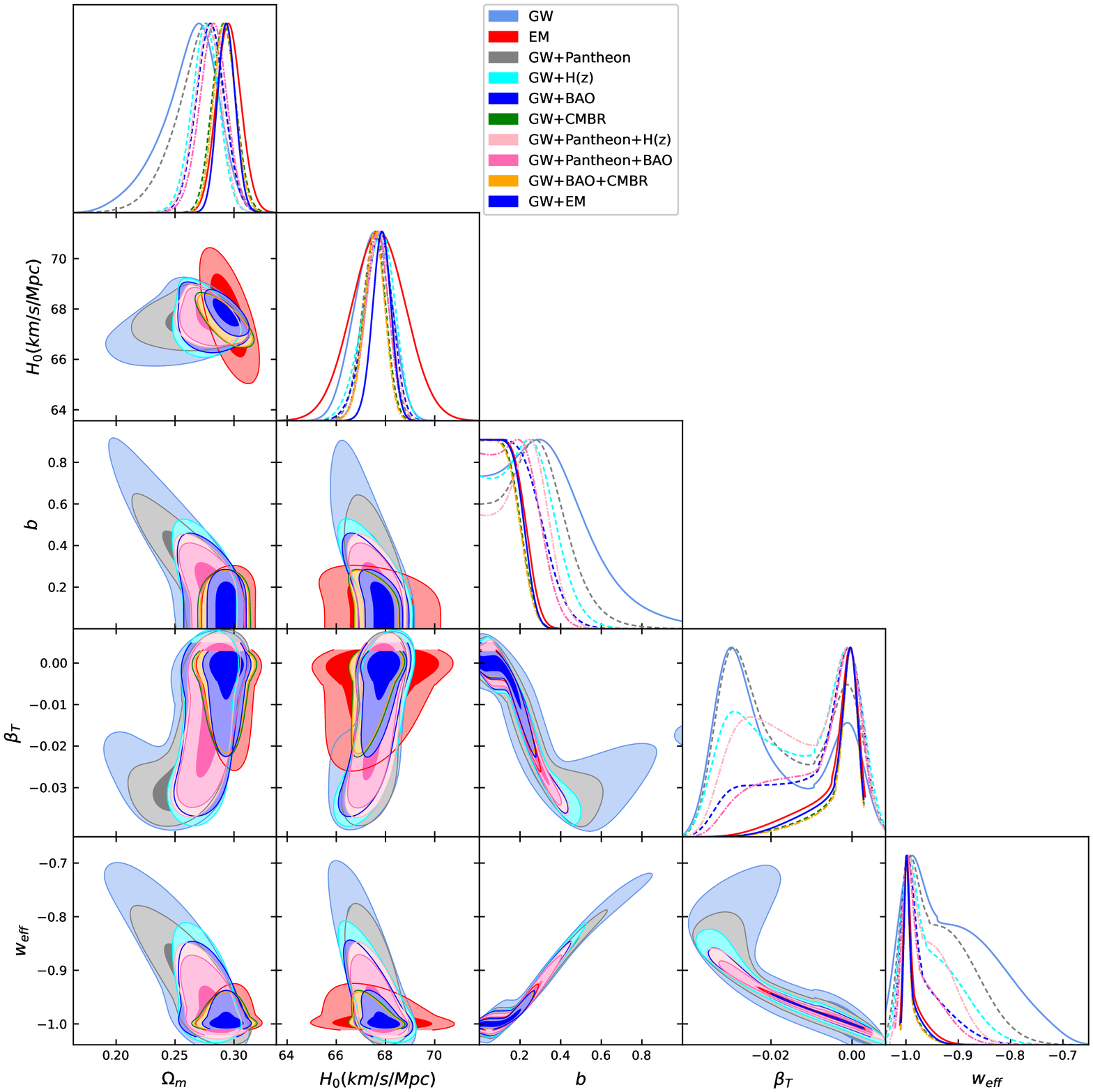}}

  \caption{ The probability density functions (pdfs) , the $68\%$  ($1\sigma$) and $95\%$ ($2\sigma$) confidence regions for the parameters  ($\Omega_{m0}$, $H_0$, $b$, $\beta_T$ and $w_eff$) in the $f_e$CDM model respectively.      }
  \label{ft2tri}
\end{figure*}

\begin{figure*} 
 \centering
{ \includegraphics[width=8cm,height=8cm]{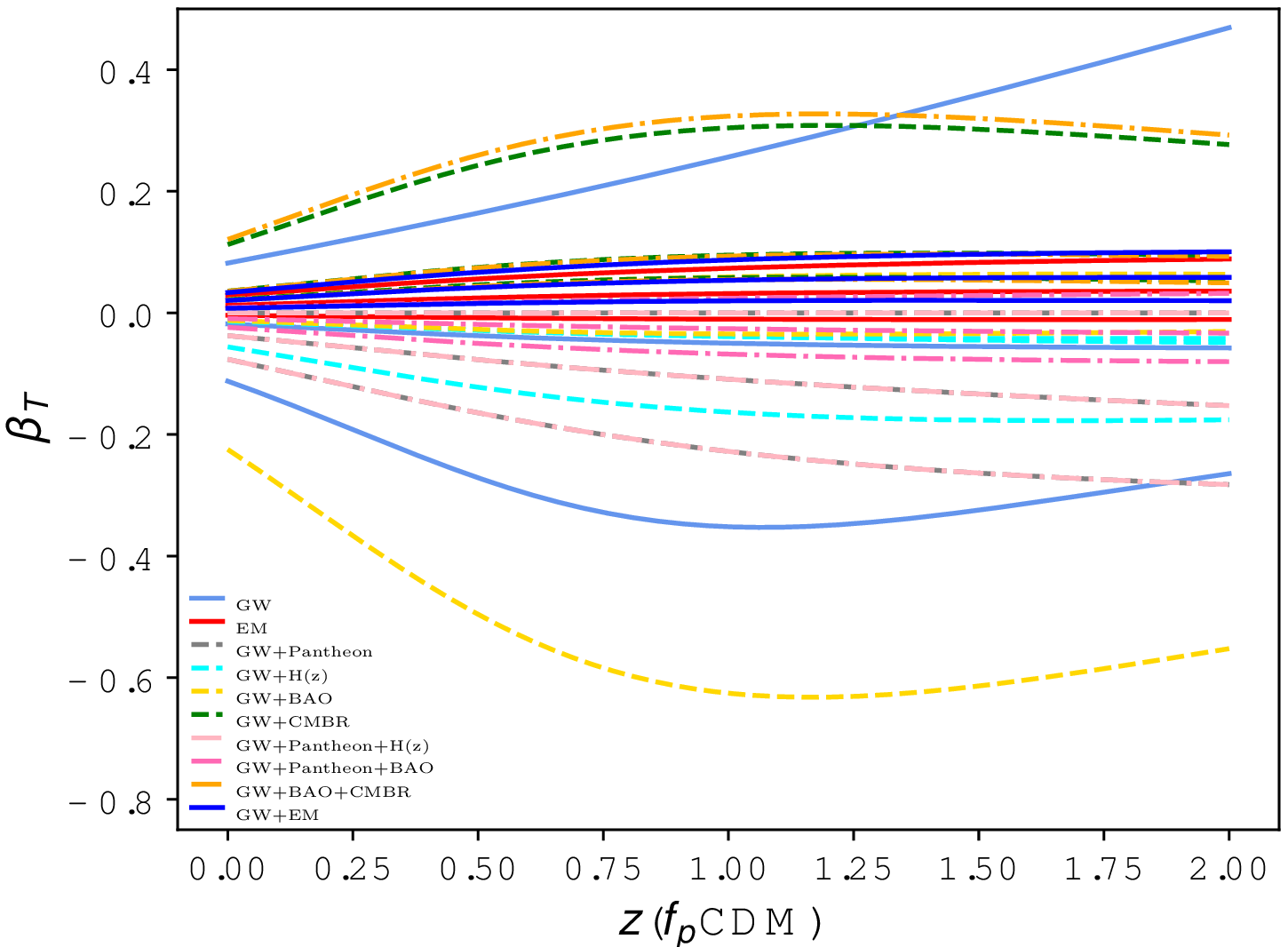}}
{ \includegraphics[width=8cm,height=8cm]{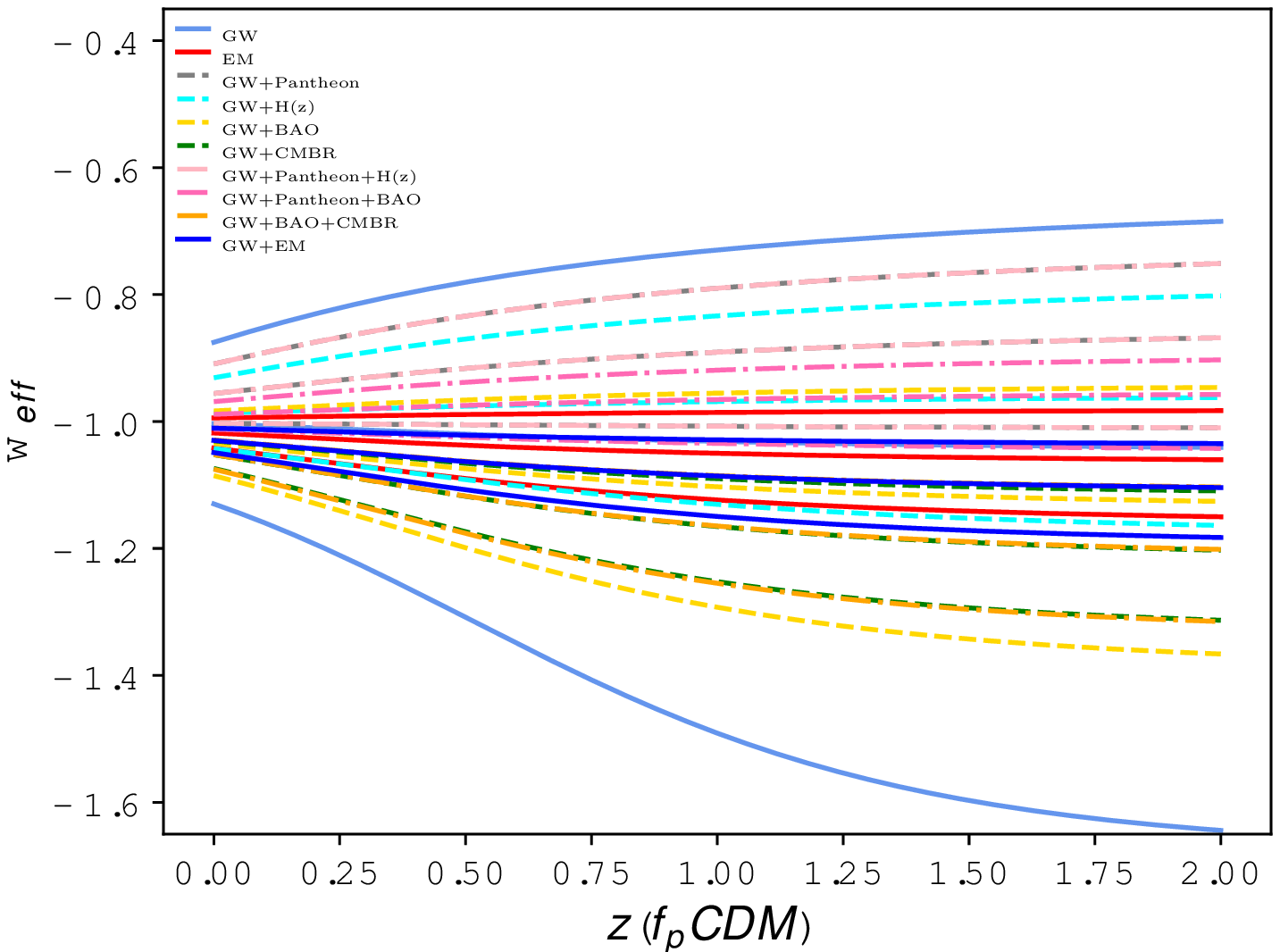}}
  \caption{The evolution of   $w_{eff}$ and  $\beta_T$  of  the  $f_p$CDM model in $1\sigma$ region for  the GW, EM and combined data separately.  }
  \label{evof1}
\end{figure*}

\begin{figure*}  
\centering
{ \includegraphics[width=8cm,height=8cm]{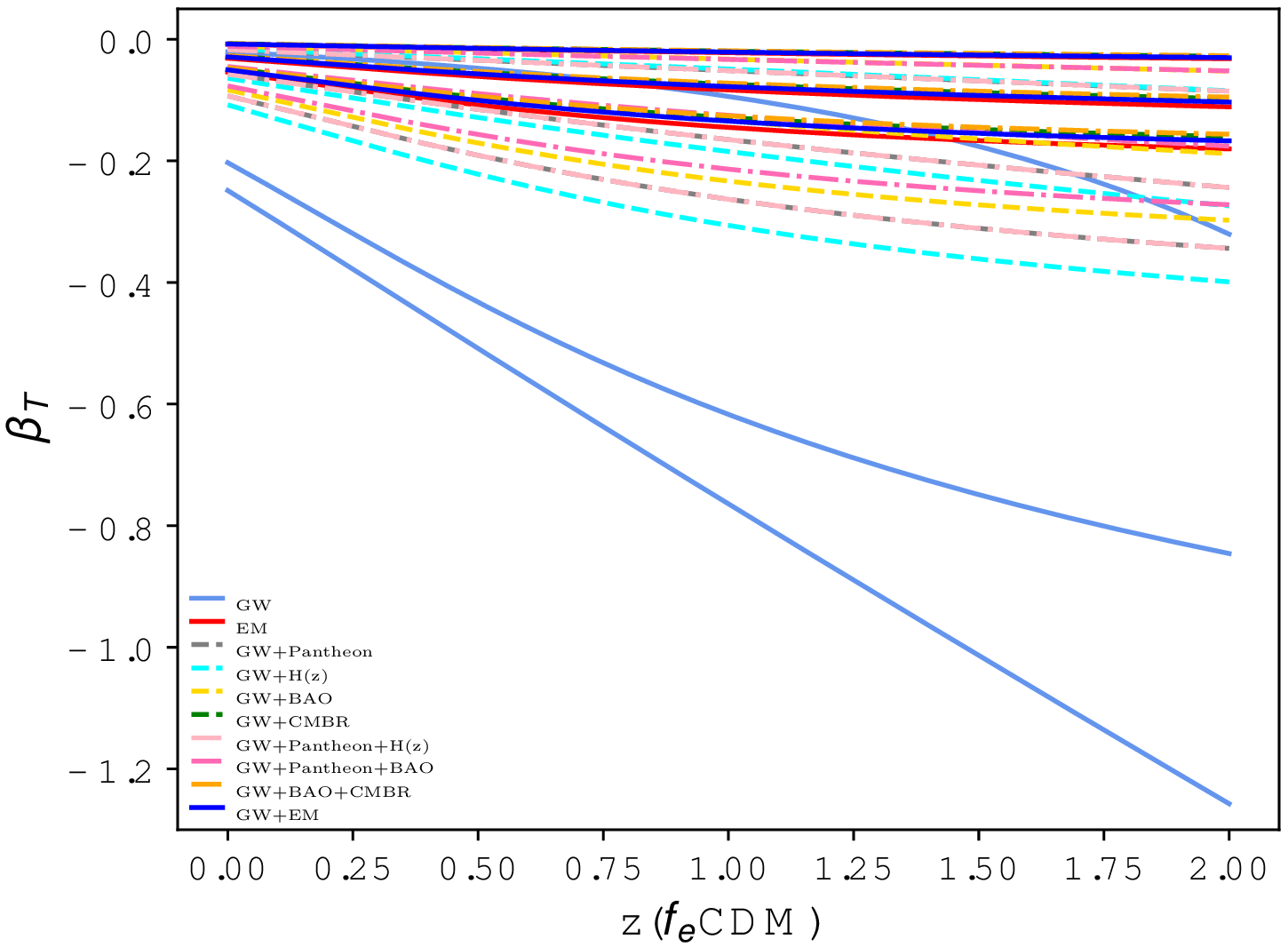}}
{ \includegraphics[width=8cm,height=8cm]{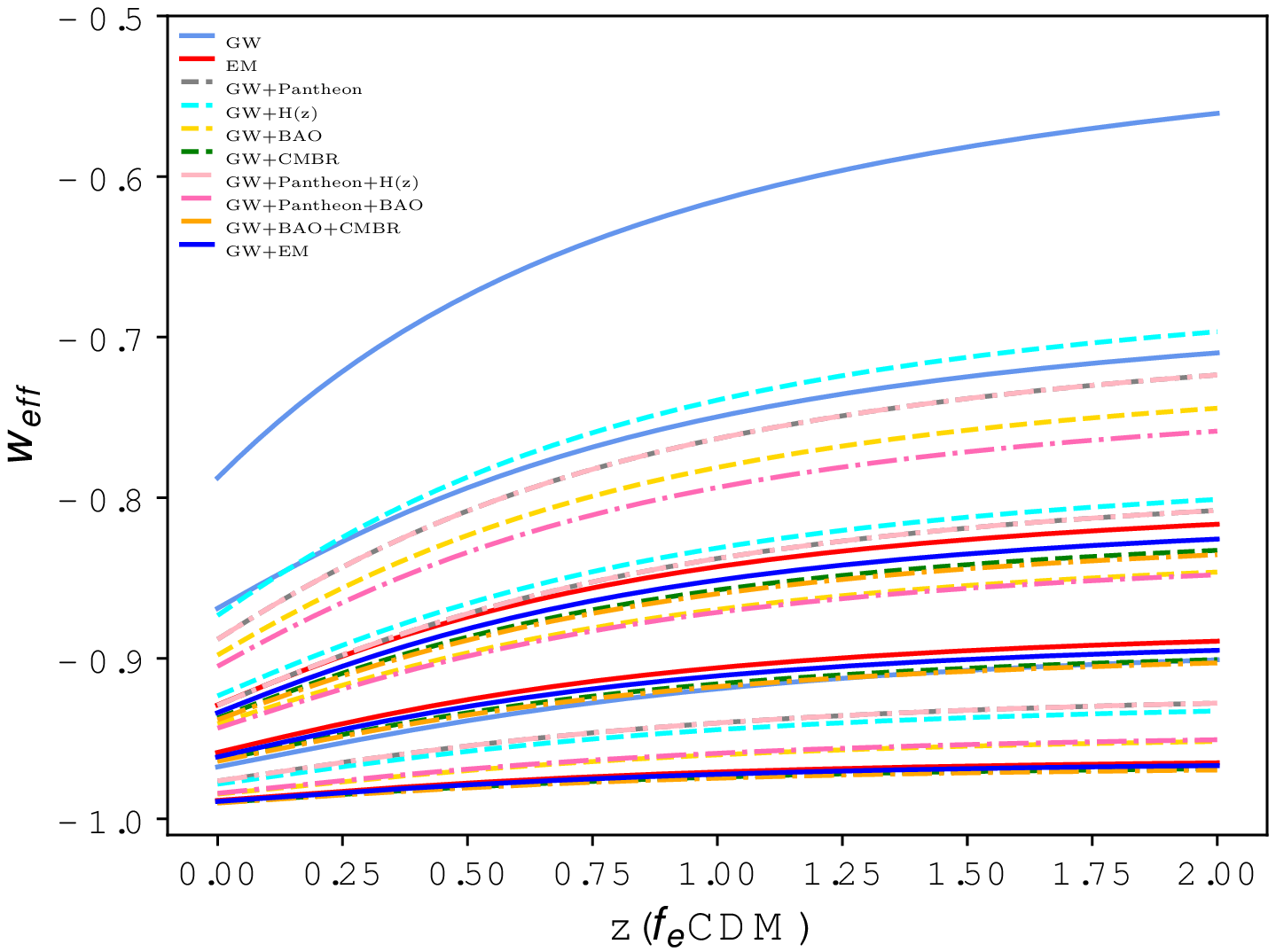}}

  \caption{The evolution of   $w_{eff}$ and  $\beta_T$  of  the  $f_e$CDM model  in $1\sigma$ region for  the GW, EM and combined data separately.   }
  \label{evof2}
\end{figure*}

\subsection{ Data Comparison}
The  Markov Chain Monte Carlo (MCMC) method \cite{Lewis:2002ah}  and the  maximum likelihood method    are used to constrain the $f(T)$ parameters. Firstly, we constrain the $f(T)$ models by using all the EM data (Pantheon$+$H(z)$+$BAO$+$CMBR). And we use the best fitted  values of the results for EM data and ET design to simulate the GW data.
  We list the simulated GW data and the  $1\sigma$ fitting results of EM data in Fig.\ref{dl}.  
  
   Comparing the best fitted value of GW data with its fiducial data, we could see  the best fitted values of parameter $\Omega_{m0}$, $b$ and $H_0$  are smaller than the  fiducial data.  Meanwhile,  $\beta_T$ is smaller the fiducial value and  $w_{eff}$ is larger than the fiducial value.
  Furthermore, the constraining regions of $\Omega_{m0}$, $b$, $\beta_T$ and $w_{eff}$ given by GW constraint are much larger than that of  the EM data. While   that of 
GW data lead to a comparative  constraint   with  the EM data for the $H_0$ parameter.  Then,  through the fiducial values are changed in the GW data,  the best fitted values of all the parameters are still in the $1\sigma$ region of the EM constraining results.
   We calculate the residues by using the $1 \sigma $ values to subtract the fiducial values.  The residues have normal distributions which denote there is no evidence for a systematic difference between  GW and  EM-based estimated.  Then, the simulated GW data  are consistent with the  EM data.  And,  the combinations of GW data with the EM data is reasonable.

 As  the $H_0$ tension is mainly  between the early data and the late  data, we consider three ways of combinations.
 Firstly, we combined  GW with the Pantheon, H(z), BAO and CMBR data separately. Then  we  combined GW to the late measurements( Pantheon $+$ H(z))  and the early measurements ( BAO
$+$CMBR) separately. And, the crossing  combination (GW$+$BAO$+$Pantheon) is added as supplements.   At last, we combined them all.
   To denote the $H_0$ problem,  the data  tension \cite{Raveri:2018wln}  is calculated
\begin{eqnarray}
T_1(\theta)=\frac{|\theta(D_1)-\theta(D_2)|}{\sqrt{\sigma_{\theta}^2(D_1)+\sigma_{\theta}^2(D_2)}}
\end{eqnarray}
where $\theta$ is chosen as the best fitted values\ of $H_0$,   the first data set  $D_1$ is separately chosen as  the  three kinds of data   discussed in the above,  the second data is set as the SH0Es  Teams data  ($H_0 = 73.2 \pm1.3 km/s/Mpc$ at $68\%$ confidence level) in this letter.

\section{Constraining results}\label{result}
In Table \ref{tab},  we show the constraining results of the constant parameters $\Omega_{m0}$,
 $H_0$, and  $b$. Furthermore,  we compute the  evolving parameters  $\beta_T$ and $w_{eff}$ as well.
 In Figs. \ref{ft1tri} and \ref{ft2tri} , we present the triangle plots of the parameters which include the probability density function (pdf)  and the  contours for  both $f(T)$ models.
 In addition, we plot the $1\sigma$ evolutions of $\beta_T$ and $w_{eff}$  in Figs. \ref{evof1} and \ref{evof2}
  \footnote{    The $\chi^2$ of each model for each data set  are equal to  the number of degree of freedom, the constraints are reasonable. Especially, the $\chi^2$ of GW data of $f_p$CDM is smaller than that of  $f_e$CDM.    This
 comparison is trustworthy because  the GW data are simulated from different models. Then we do not list the $\chi^2$ values in this letter.}.

     \subsection{The $f_p$CDM model}

     The contours are closed and    smooth and the pdfs are gaussian-distributed in $f_p$CDM model. 
    Explicitly,  either the EM or GW data alone can not distinguish the $f(T)$ model from the $\Lambda$CDM model.
The constraining tendencies of GW$+$Pantheon   and GW$+$H(z) data are similar which give out the positive best fitted $b$ and negative best fitted $\beta_T$.  In the contrast, that of  GW$+$BAO   and GW$+$CMBR  data  are similar which give the negative best fitted $b$ and positive best fitted $\beta_T$.  This is  caused by the  $H_0$ tension between the early and late EM data, and denotes the parameters $\beta_T$ and $b$ negatively related.

 Because of the $H_0$ tension,   the GW$+$Pantheon$+$H(z) and GW$+$BAO$+$CMBR both have tight constraining results.
    And   the GW$+$BAO$+$CMBR data has the tightest constrain while  the GW$+$EM one which includes all the  EM data does not.
 The GW$+$Pantheon$+$BAO data has larger contours than the   GW
 $+$Pantheon$+$H(z) and GW$+$BAO$+$CMBR data which denotes the $H_0$ tension as well.     Obviously,  the GW data combined with both the early and late  EM  data improve the constraint significantly.
 After the solution of the  $H_0$ tension,  the distinguishable problem will be more clear.

   The $w_{eff}=0$ value is not included in $2\sigma$ regime for $f_p$CDM model by the constraining of the GW$+$BAO$+$CMBR data. In another saying,
 the  $f_p$CDM model could be distinguished from $\Lambda$CDM  in $2\sigma$ constraints by adding the GW data.
 The distinguishable effect  is mainly from the   $\beta_T$ parameter which is related to the  GW luminosity data and breaks the main degeneration between the parameters.


\subsection{The $f_e$CDM model}

   As shown in Table \ref{tab},  we obtain  a minus $\beta_T$ and a quintessence-like $w_{eff}$ in $f_e$CDM model as the setting of the prior  $b>0$.
     The evolutions of $w_{eff}$ and $\beta_T$   almost follow $\Lambda$CDM model at late time which are quite flat.
     The contours  of $\beta_T$ have two peaks for $f_e$CDM model.   There is a  non-gaussian behavior  for $b$ which is caused by the prior $b>0$.
Though the contours  related to  $b$ are not closed, but the  contours related to  $w_{eff}$ and $\beta_T$   are nearly closed.

Generally,  the  GW$+$Pantheon data constrains the $b$ parameter more effectively in the $f_e$CDM model. And, the best fitted value of  all the  data of the $\beta_T$  and $w_{eff}$  are close to $\Lambda$CDM model.  As   we set the $b>0$ prior for $f_e$CDM model,  it does not face to the distinguishable problem.

\subsection{ Short summary}

Comparing the constraining results for $f_p$CDM and $f_e$CDM models,  the $1\sigma$ constraining regions of all the  GW related data  of $f_e$CDM model are comparable with  that of $f_p$CDM model. The  constraining precisions do not  sensitive to the prior $b>0$.  And the shape of the $f(T)$ function  almost does not affect the constraining precision as well.
And considering the tension between our data and the SH0Es data as list in Table \ref{tab},    the combined  data have larger tensions than the GW or EM data, that is mainly because the precision of the data are improved, while the shift of the best-fitted value is slight.
 
Finally, as shown in Figs. \ref{evof1} and  \ref{evof2}, at high $z$, the deviations from $\Lambda$CDM model become evident.
Then more high $z$ data may help to
   distinguish them.

\section{ Conclusion  }\label{sum}

In this paper, we focused on the detection ability of the future GW data to constrain the $f(T)$ models and discussed  the related  $H_0$ tension problem.  The GW$+$BAO$+$CMBR data give the tightest constraints.  The GW data are powerful probes to distinguish the modified gravity from the GR, especially the degenerated ones in view of early EM data. As our computations show, the gravitational wave detections offer a remarkable approach to explore the universe from a new perspective, providing an access to astrophysical processes that are completely ignorant to EM observations.

 {\bf Acknowledgments.}
  YZ thanks  Dr. Jingzhao Qi, Prof. Hao Wei, Dr.  Tao Yang, Prof.Wen Zhao,
YZ is supported by National Natural Science Foundation of China under Grant No.11905023, CQ CSTC under grant
cstc2020jcyj-msxmX0810 and cstc2020jcyj-msxmX0555. HZ is supported by Shandong Province Natural Science Foundation under grant No.  ZR201709220395, and the National Key Research and Development Program of China (No. 2020YFC2201400).

\end{document}